\documentclass[preprint2]{aastex}

\shorttitle{Space weathering of basaltic asteroids}
\shortauthors{Marchi et al}

\begin{document}

\newcommand{\mum}{$\mu$m}

\title{On the puzzle of space weathering alteration of basaltic asteroids}


\author{S. Marchi}
\affil{Department of Astronomy, Padova University \\ 
Vicolo dell'Osservatorio 3, I-35122 Padova, Italy}
\email{simone.marchi@unipd.it}

\author{M.~C.  De Sanctis}  
\affil{Istituto di  Astrofisica  Spaziale e Fisica  Cosmica, INAF\\  
Via  Fosso del  Cavaliere  100, I-00133  Roma,  Italy}
\email{mariacristina.desanctis@iasf-roma.inaf.it}

\author{M. Lazzarin}
\affil{Department of Astronomy, Padova University \\ 
Vicolo dell'Osservatorio 3, I-35122 Padova, Italy}
\email{monica.lazzarin@unipd.it}

\author{S. Magrin}
\affil{Department of Astronomy, Padova University \\ 
Vicolo dell'Osservatorio 3, I-35122 Padova, Italy}
\email{sara.magrin.1@unipd.it}

\begin{abstract}

The majority of  basaltic asteroids are found in  the inner main belt,
although a  few have  also been  observed in the  outer main  belt and
near-Earth  space.   These asteroids  -referred  to  as V-types-  have
surface compositions  that resemble that of the  530~km sized asteroid
Vesta. Besides  the compositional similarity,  dynamical evidence also
links many V-type  asteroids to Vesta.  Moreover, Vesta  is one of the
few  asteroids to  have been  identified as  source of  specific
classes of  meteorites, the howardite,  eucrite, diogenite achondrites
(HEDs).\\
Despite  the  general  consensus  on the  outlined  scenario,  several
questions remain  unresolved.  In particular,  it is not clear  if the
observed spectral  diversity among Vesta,  V-types and HEDs is  due to
space weathering, as is thought to be the case for S-type asteroids.\\
In  this paper, SDSS  photometry is  used to  address the  question of
whether the spectral diversity among candidate V-types and HEDs can be
explained by  space weathering.  We show that  visible spectral slopes
of  V-types are  systematically  redder  with respect  to  HEDs, in  a
similar way  to what  is found for  ordinary chondrite  meteorites and
S-types.  On  the assumption that space weathering  is responsible for
the  slope mismatch,  we estimated  an upper  limit for  the reddening
timescale of about 0.5~Ga.   Nevertheless, the observed slope mismatch
between HEDs  and V-types poses  several puzzles to  understanding its
origin.  The  implication of  our findings is also  discussed  in the
light of Dawn mission to Vesta.

\end{abstract}


\section{Introduction}

Vesta is  the largest basaltic fully  differentiated asteroid.  Remote
visible  and  near-infrared  spectroscopy  indicates the  presence  of
basaltic mineralogy  on Vesta's surface  and the possible  presence of
other components  \citep{mcc70,lar75,gaf97}.  A large  number of small
asteroids  that  show a  similar  surface  composition, the  so-called
V-type asteroids,  populates the inner  main belt \citep{bin93,duf04}.
V-type asteroids have  dimensions less than a few  tens of km that cannot
sustain  differentiation  processes,  therefore  they must  have  been
originated  from  one -or  more-  much  larger, differentiated  parent
body. Thus, it has been suggested  that Vesta is the parent body for
V-types \citep{mcc70}.\\
To support  this scenario, apart from  the aforementioned compositional
similarity, it has been shown that  many V-types and Vesta belong to a
large  asteroid  family  \citep[][]{mar96,nes08}.  To  reinforce  this
link, a 460~km wide crater has been detected on Vesta \citep{tho97}.\\
Another interesting aspect concerning  Vesta and V-types is that their
visible and  near-infrared spectra  have been found  to be  similar to
those  of a  particular  suite of  meteorites,  namely the  howardite,
eucrite              and             diogenite             achondrites
\citep[HEDs;][]{mcc70,con77,tak97,dra01}.\\
On the other hand, it is  also clear that neither all V-types nor HEDs
are  compatible  with  being  derived  from Vesta;  therefore,  it  is
suspected  that  they originate  from  multiple differentiated  parent
bodies \citep{yam02,wie04}.
In  this respect,  it  is significant  that  a few  V-types have  been
discovered  beyond the 3:1  mean motion  resonance \citep{laz00,duf09}
which   cannot    be   related    to   Vesta   on    dynamical   basis
\citep{roi08,nes08}  and -at  least in  the  case of  Magnya- also  on
compositional basis \citep{har04}.\\
Despite  the general  scenario  described above  is largely  accepted,
there  are some observations  that remain  unexplained. First  of all,
V-types  generally exhibit  stronger absorption  bands than  Vesta and
steeper  continuum  both  in  the  visible \citep{hir98}  and  in  the
0.4-1.6~\mum\ region \citep{bur01}.\\
The reason for these discrepancies  is not yet clear.  Possible causes
are indicated in composition/texture differences or due to the effects
of  some alteration process  \citep[e.g.][]{bur01}, like  for instance
those  that have  been  demonstrated to  cause  optical alteration  on
silicate-rich  S-type  asteroids  \citep{jed04,mar06a}.   However,  on
S-types the 1 and 2~\mum\ absorption bands tend to become shallower by
increasing space  weathering (i.e.  increasing  spectral slope), while
this effect  is not observed  on V-types.  V-type asteroids  that have
reddest  slopes  also  show  strong  1 and  2~\mum\  absorption  bands
\citep[e.g.][]{bur01}.
Vesta is found to be most comparable to howardites with a fine-grained
size distribution \citep[e.g.][]{pie06}.\\
On the  other hand, the  laboratory experiments conducted on  HEDs and
pyroxenes confirm that V-type materials should alter under the effects
of ion  bombardment and  micrometeorite impacts, in  a way  similar to
S-types, although the alteration  timescale is likely somehow reduced
\citep{hir98,mar05a,ver06}.\\
In this paper we present  a comprehensive model of space weathering of
candidate  V-types detected  via SDSS  photometry  \citep{roi06}.  The
present analysis  aims to investigate  whether or not  V-type surfaces
are  affected  by space  weathering  processes.   We identify  several
interesting  facts  that  support  the presence  of  space  weathering
alteration, although  its behavior appears  distinct to what  found on
S-types.\\
 In  doing so, our  findings disclose  interesting aspects  of surface
 properties of Vesta  and V-types that are useful  also in the context
 of  the NASA  Dawn mission  \citep{rus07} for  what  concerns mission
 planning and Vesta's data interpretation.

\section{Modeling the space weathering of V-type asteroids}

\subsubsubsection{Spectrophotometric catalog of V-types}

It has  been demonstrated for  S-type asteroids that  visible spectral
slopes can  be successfully used  to model the optical  alterations of
asteroidal surfaces due to space weathering. Spectral slopes have been
evaluated   using   different   methods,   mainly   photometric   data
\citep[e.g.][]{jed04}  and spectroscopy  \citep{mar06a,pao07}, leading
basically to the same conclusions.\\
Concerning  V-types,  spectroscopic  data  are available  only  for  a
restricted  number ($\sim$70) of  objects, preventing  any significant
characterization  of the whole  population.  For  this reason,  in the
present  work, we will  make use  of SDSS  photometry reported  in the
Moving Object  Catalog V3 \citep[MOCV3;][]{ive01,jur02},  which allows
the analysis to be extended to  include a large number of objects (388
and 466  family and  non-family V-types, respectively).   V-types have
been selected according to  \cite{roi06}. In addition, we also include
52 V-types  identified by the spectroscopic surveys  SMASS, S3OS2, and
SINEO \citep{laz04,bus02,bin04,mar05b}  which are not  included in the
SDSS MOCV3. It is important  to underline that the list established by
\cite{roi06}   consists   of   {\it   candidate}  V-types   and   that
spectroscopic   data   are   needed   for  a   robust   identification.
Nevertheless,  a comparison  between candidate  V-types  and confirmed
ones displays a good match \citep{roi06}.\\
Spectral slopes have  been evaluated by linear best  fit of the albedo
using  the SDSS  bands  g', r',i',  z'.  One sigma  errors of  SDSS
  albedoes  have  been  used  as  weighting factors  for  the  fitting
  procedure.   As  for spectroscopic  data,  we  first evaluated  the
albedo in g', r', i', z' bands through cubic spline of the spectra and
then proceeded with linear best  fit.  Albedoes have been normalized to
the r' band \citep{roi06}.

\subsubsubsection{Parameterizing the space weathering}

Previous  analysis of  S-types and  ordinary chrondrites  (OCs) showed
that the space weathering on asteroids  is a process that, if no other
alteration is  present, is expected to progressively  evolve over time
until  eventually a  sort  of saturation  is  reached.  Moreover,  its
efficiency depends on  the location and past orbital  evolution of the
asteroids.\\
In this work,  the study of space weathering is  done according to the
model  developed in  a series  of papers  by  \cite{mar06a,mar06b} and
\cite{pao07}.  This  model demonstrated that the  aging, location, and
past  evolution  can  be  condensed  into a  single  parameter,  named
exposure  $E$.    The  exposure  is   proportional  to  the   dose  of
radiation/ion received from the Sun and scales as

\begin{equation}
E=\frac{T}{a^2\sqrt{1-e^2}}
\end{equation}

where  $a, e$  are the  Keplerian elements  of the  body, and  $T$ is its
average  collisional  age \citep{mar06a}.   For  MBAs  we used  proper
elements  given their  stability  on  a longer  timescale. Ages  are
derived according to the  collisional evolution of main belt asteroids
\citep{bot05,mar06a}.    Collisional   lifetimes   depend   upon   the
diameters,  which  have  been  estimated using  the  measured  average
geometric     albedo    (0.3)     of    a     sample     of    V-types
\citep{ted02,ben02,del03,del06}.  The collisional age of V-types spans
from about 0.5  to 2.7~Ga.  These extremes correspond  to the smallest
and largest  bodies of the  sample, which have estimated  diameters of
about 1~km  and 10~km,  respectively. In what  follows, it  is assumed
that  the formation  of Vesta's  family occurred  sometime  before the
oldest age in the sample,  namely $\sim2.7$~Ga, so that asteroid's age
are set by subsequent  collisional evolution.  Although Vesta's family
age  is not  well constrained,  a  recent work  of \cite{nes08}  found
evidence for  a much earlier  origin, possibly dating  3.5-3.8~Ga ago,
than previously assumed \citep{mar96}.\\
%
%
For  the NEOs,  given their  short evolution  timescale  in near-Earth
space ($\sim5$~Myr), the major  contribution to their space weathering
alteration   comes   from  their   past   evolution   into  the   main
belt. Therefore, for each NEO we set $a$ equal to its computed average
source distance \citep{mar06a} and $e=0.17$ (the average of MBAs).

\section{Space weathering on V-types?}

The main results of our model can be investigated with the help of the
slope-exposure scatter  plot.  For this purpose,  we separated V-types
belonging  to  the  Vesta  dynamical family  from  non-family  members
\citep{roi06}.  Results are shown in Figure \ref{f1}.  For comparison,
the trend derived for S-types is  also shown.  Note that the latter is
computed by the least-square fitting of S-type visible spectra of MBAs
detected by  SMASS and S3OS2 surveys  (a total of  750 objects), using
the SDSS bands  fitting procedure, as described above.   This trend is
very  close to  the  one derived  by  means of  spectroscopic data  in
\cite{pao07} (valid  for $E>10$), ensuring that  the procedure adopted
here is robust.
Figure \ref{f1} also reports V-type  NEOs along with average slopes and
range  of slopes  for a  sample of  50 HEDs.   Note that  these HED
samples  have  similar grain  size  ($<25$~\mum);  thus, the  observed
spread in  slope is likely  due to differences in  composition, rather
than due to grain size effects.\\
A number of  interesting results can be derived  from Figure \ref{f1}.
First  of  all, V-types  are,  on  average,  redder than  HEDs.   This
difference is  minimized for eucrite meteorites. To  better study this
difference, we  report the slope  histograms in Figure  \ref{f2}.  The
average  slope   gap  between  eucrites  and  MBA   V-types  is  about
0.6~\mum$^{-1}$,  similar to  that of  S-types and  OCs \citep{pao07}.
The  largest  difference,  of  about  1.0~\mum$^{-1}$,  is  found  for
diogenites and MBA V-types.\\
It is  not easy  to understand the  origin for this  marked difference
with available  data.  It  could be due,  for instance,  to systematic
differences  in grain size  or observational  conditions, such  as the
phase angle.  However, it is hard to believe that these factors can be
entirely  responsible  for  the  observed  slope  variations  for  the
following reasons.  Concerning the  grain size, we computed the effect
of  the particle  size on  the  slope by  analyzing two  eucrites and  one
diogenite  for which  spectra with  different particle  sizes  (in the
range  25-250~\mum)  have been  gathered  (RELAB  database).  In  this
range,  the slope increases  with decreasing  particle sizes  by about
0.4~\mum$^{-1}$ \citep[see also][]{hir98,bur01}.  Therefore, to obtain
the  observed reddening on  asteroids, their  particle size  should be
much smaller than 25~\mum,  a conclusion which seems unrealistic given
the low-gravity environment  of V-types\footnote{For  comparison, the
  average  regolith particle  size on  the Moon  is 70~\mum,  and only
  10-20\% of  the particles have sizes  $<20$~\mum.  Moreover, thermal
  inertia modeling shows that asteroid's regolith is much coarser than
  the lunar one \citep[e.g.][]{del09}.}.  Moreover, we recall that the
HED  slopes  used here  are  virtually  the  reddest, given  that  the
particle size of these samples is below 25~\mum.  Concerning the phase
angle of observations,  it spans from about 2  to 30~deg, which is
not too different from  geometry conditions during spectra acquisition
of  HED samples  \citep{bur01}.   There also  seem  to be  differences
between  family and  non-family V-types.   The latter  shows  a larger
slope scatter  and has a  considerable number of objects  having slope
$<-0.6$~\mum$^{-1}$  and   $>0.5$~\mum$^{-1}$  (see  Fig.   \ref{f2}).
These two extreme  classes of object are separated by  the bulk of the
slope distribution,  and might  represent two distinct  populations of
objects.   The ``blue''  non-family V-types  ($<-0.6$~\mum$^{-1}$) are
compatible  with  HEDs,  and  do  not  show  any  effect  of  spectral
alteration.\\
The ``red''  V-types ($>0.5$~\mum$^{-1}$), are much  redder than HEDs.
Their  exposure is  not systematically  different than  other V-types;
therefore the observed red slope  is not justifiable in terms of space
weathering.  Possible interpretations are that:  either i) they are S-types
that are  erroneously classified as  V-types or ii) if  basaltic, they
have a  peculiar composition  which is not  sampled by HEDs.   After a
close look at  SDSS data for these objects, it seems  that i) could be
ruled  out;  however,  v-nir  spectroscopy  is  needed  for  a  secure
classification (see end  of this section for further  comments on this
issue).   Moreover,  these  extreme  slope  objects do  not  have  any
significant difference in terms  of orbital parameters with respect to
other V-types.  Although  less evident, blue and red  objects are also
present among family V-types.\\
Interestingly, NEO V-types are somehow  in between HEDs and MBAs. This
is,  at  least  for  some  NEOs,  in  agreement  with  their  computed
intermediate  exposure,  as  also  found  for  S-types  \citep{pao07}.
Nevertheless, the optical properties of  NEOs can be affected by other
processes like  close encounters \citep{mar06b,bin10}.  The effects of
the latter on V-type NEOs will be investigated in a future work.\\
%
%
Alternatively, several differences with S-types can be noted. The most
striking  is that  the slope  shows  a negative  trend for  increasing
exposures.  In  other words, the  more exposed objects tend  to appear
bluer,  on  average,  than  less  exposed ones.   The  trend  is  less
pronounced for  non-family members, and  becomes remarkably noticeable
for family  members.  A linear  best fit performed  through minimizing
chi-squared error statistics gives  the following trends (expressed in
units    of     slope[\mum$^{-1}$]/$\log(E[{\rm    Ma    AU^{-2}}])$):
$-0.71\pm0.02$ and  $-0.35\pm0.02$ for family  and non-family V-types,
respectively.  Error  bars are  estimated taking into  account 1-sigma
errors  of  slope  values.    Both  trends  are  statistically  highly
significant   with   a   t-distribution  two-tailed   probability   of
significance  of   $10^{-6}$  and  $10^{-9}$,  respectively\footnote{A
  two-tailed   probability   less   than  $10^{-2}$   indicates   full
  statistical significance of the correlation.}.  For comparison, the
S-types   trend   is    $0.23\pm0.07$   (two-tailed   probability   of
$10^{-3}$).\\
%
%
In   light of the above  results, the unexplained  lack of spectral
reddening on  Vesta can be studied  from a new  perspective.  First of
all, there  seems to be  a general de-reddening trend  with increasing
exposures.   These trends  are  also marginally  present  in terms  of
diameters, and  become progressively more pronounced in  terms of ages
and exposures.   This result suggests  that the trends are  not merely
due to  an asteroid size effect.   Note that Vesta  being merely the
largest body (i.e. the most exposed) fits with this trend.\\
%
%
The slope scatter of V-types is  larger than that of HEDs, however the
latter may be limited due to the relatively low number of samples used
in this  work.  Therefore, as for  the HEDs, it  is possible that
  much of the V-type slope spread may be due to compositional/texture
  differences.  It  could also be   (partially) due to the  presence of
  S-type -erroneously  classified as V-types-  interlopers within the
  lists used in this work.   S-types are expected to have, on average,
  higher  slopes than V-types  \citep{bus02}. Therefore,  the possible
  presence of S-type interlopers may affect the observed excess of red
  objects, which likely has a negligible influence on the clear slope
  gap between HED and candidate V-types shown in Fig.  \ref{f2}.\\
%

\section{Discussion and Conclusion}

\begin{figure*}[h]
\includegraphics[angle=-90,width=15cm]{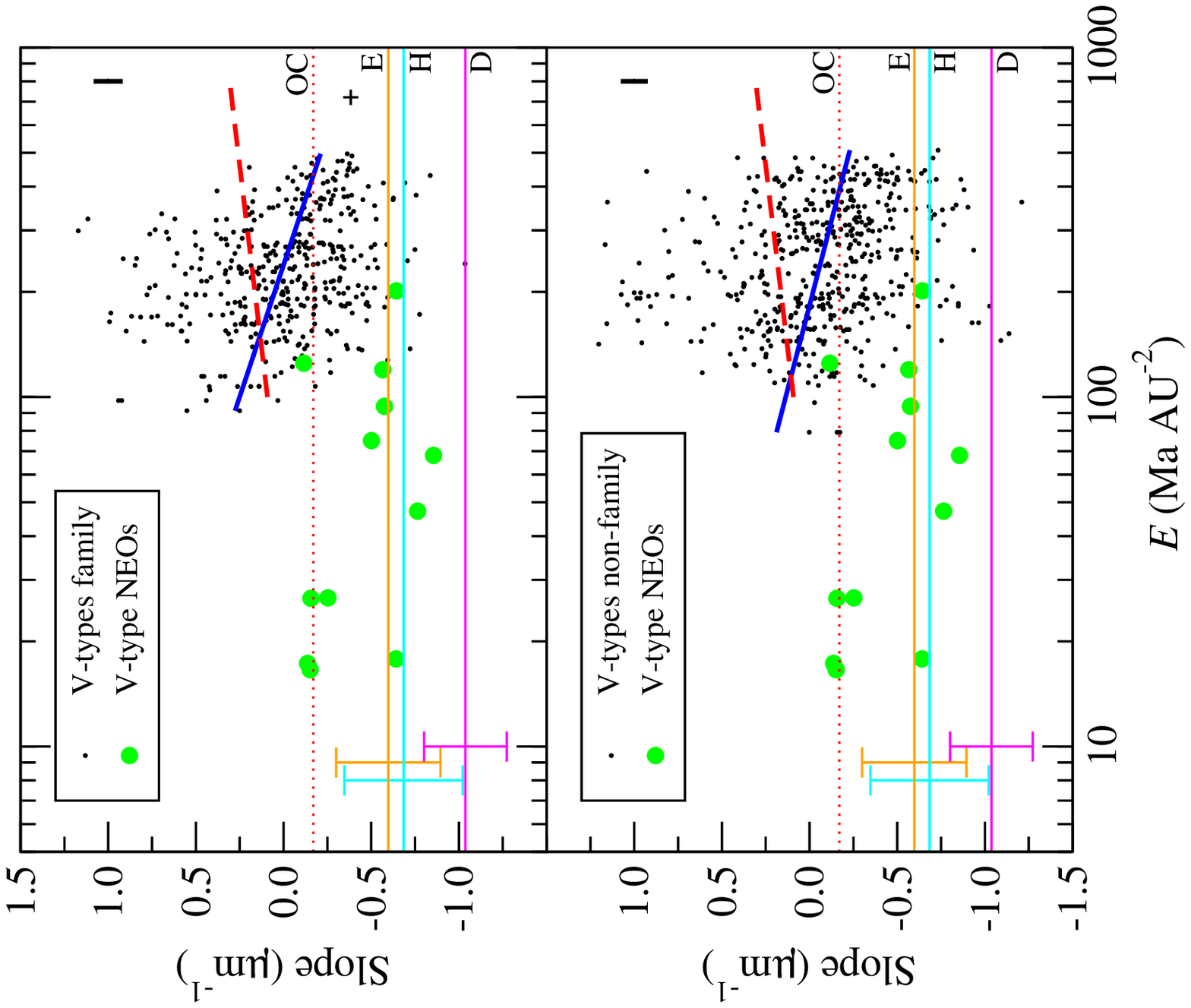}   
\vspace{0.1cm}
\caption{Slope vs  exposure scatter plot for MBA  V-types. Both family
  and  non-family V-types  are  reported.  The  position  of Vesta  is
  indicated by the plus sign (upper panel).  Horizontal lines indicate
  the average  values for a sample  of 14 howardites,  26 eucrites, 10
  diogenites and 180 OCs. The range of slope values for HEDs (computed
  as 1-sigma of  the slopes) is also reported  (placed arbitrarily at
  $E=8,9,10$).   NEOs slopes  are indicated  by filled  green circles.
  Red  dashed line indicates  the reddening  trend for  S-types, while
  blue lines indicate  MBA V-types best fit. The average 1-sigma
    slope  error  is  reported  at  the  upper-right  corner  of  both
    panels.}
\label{f1}
\end{figure*}

\begin{figure*}[h]
\includegraphics[angle=-90,width=15cm]{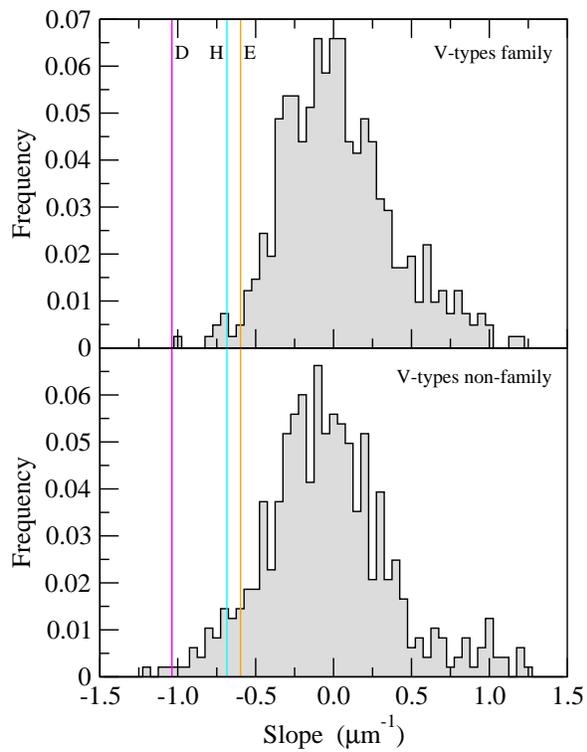}   
\vspace{0.1cm}
\caption{Histograms  of  visible  slopes  for  family  and  non-family
  V-types.   Average  slope  values  of  HEDs are  also  indicated  by
  vertical  lines   (see  also  Figure  \ref{f1}).    Note  that  both
  distributions show  marked non-Gaussian shapes, with  a clear excess
  of    ``blue''   and   ``red''    objects,   i.e.    having   slopes
  $<-0.6$~\mum$^{-1}$  and $>0.5$~\mum$^{-1}$, respectively  (see the text
  for details). }
\label{f2}
\end{figure*}

The main result of this paper is that V-type slopes are systematically
redder than  HEDs (see Fig.   \ref{f1}; Fig.  \ref{f2}).  Thanks  to a
large  sample  of candidate  V-types,  the  present work  considerably
extends  previous  results  \citep[e.g.][]{hir98,  bur01}  and  allows
quantitative  estimates of  the  reddening of  V-types.  The  spectral
mismatch mimics what was found for OCs and S-types \citep{pao07}.  The
observed  discrepancy  could be  due  to  factors  such as  systematic
compositional/texture    differences,    different    conditions    of
observations  or  space  weathering.   We  showed  that  observational
conditions and texture  are unlikely to be the  source of the observed
discrepancy.  Therefore,  either the  marked slope shift  between HEDs
and V-types is  due to systematical differences in  composition, or it
can  be ascribed  to some  kind of  space weathering  alteration.  The
previous hypothesis would imply that  the suite of HED compositions is
not  representative of  the whole  V-type population.   While  this is
certainly an open possibility, it is mandatory to obtain v-nir spectra
of small V-types  to confirm.  In the case that  the slope mismatch is
due to  space weathering  processes, it is  possible to  constrain the
reddening timescale on the basis  of the collisional ages.  Although a
precise estimate is not possible with present data, an upper limit for
the reddening timescale is set  by the youngest V-type MBAs which have
ages  of $\sim$0.5~Ga.   Moreover, 58\%  of V-type  NEOs have  a slope
compatible with  HEDs (within 1-sigma  of average values),  while only
about    20\%   of    S-type    NEOs   are    compatible   with    OCs
\citep{bin10}. Therefore, if  no other effect comes into  play, it can
be inferred that V-type's space  weathering timescale is larger than a
factor of $\sim$3 with respect to that of S-types.\\
The possibility that space  weathering is responsible for the spectral
mismatch is also  reinforced by the few experiments  published to date
of  ion   bombardment  of  Bereba  eucrite   and  Johnstown  diogenite
\citep{hir98,ver06}.   They both  exhibit  reddening (and  darkening).
Note  that these experiments  also confirm  a slower  timescale with
respect to OCs.\\
The  true nature  of the  spectral mismatch  still remains  elusive in
several  aspects; nevertheless  some  noteworthy characteristics  have
been identified.  In particular,  we explore in detail the possibility
that the spectral mismatch is due to space weathering processes.\\
First  of all,  the spectral  slope anti-correlates  with  exposure to
solar wind.   This is the opposite  of what was found  for S-types and
other  spectral  types  \citep{laz06}.   Therefore, if  some  kind  of
Sun-related  space  weathering  is  operating on  V-types,  either  it
behaves in a different manner with respect to S-types or other factors
wipe out  the reddening-exposure  relation.  Concerning the  latter, a
possibility is  represented by a composition  gradient across V-types,
from   eucrite-like   for  the   smallest   members  ($\sim$1~km)   to
diogenite-like for  the largest members  ($\sim$10~km).  This putative
composition gradient would help to explain the observed slope-exposure
trend  for two  reasons: i)  diogenites  are bluer,  on average,  than
eucrites;  and ii)  according to  the few  experiments  available, the
reddening  of  diogenites  seem  less pronounced  than  for  eucrites,
although  the  space parameters  of  HED  compositions and  alteration
processes are far from being exhaustively investigated. \\
Note that the  above explanation is preferred with  respect to other
possibilities,  e.g.  grain  size  variation, since  the latter  would
require a  grain size finer  than $\sim25$~\mum\ (i.e.   higher slope)
for smaller objects.  This is  the opposite of what is expected, since
smaller  objects tend  to have  a  larger grain  size due  to the  low
gravity.  Also,  the substantial presence of  S-type interlopers among
the red V-types would  not greatly affect the observed slope-exposure
trend since they  are nearly equally distributed in  terms of $E$ (see
Fig. \ref{f1}).\\
If confirmed, the above compositional trend among family V-types could
be the result of cratering ejecta from various depths in Vesta's crust
(eucrite is expected  in the upper crust, while  diogenite is expected
in the lower crust); or may  be due to different episodes of cratering
occurring  in  regions with  different  compositions.  However,  since
large fragments  are expected to  originate in the  near-surface spall
region, the  second hypothesis seems  more probable. This may also be 
confirmed  by  the  variety  $^{39}$Ar-$^{40}$Ar shock  ages  of  HEDs
\citep{bog03}.   Furthermore, the  spectral data  of V-types  seem to
indicate  that both  mineralogies are  present \citep{duf04},  but the
limited sample prevents a firm conclusion.\\
The proposed scenario could be tested as soon as more v-nir spectra of
small  V-types   would  become  available.   Note   that  the  above
discussion  holds for  both family  and non-family  V-types.   For the
latter,  however,  the anti-correlation  is  less  pronounced, and  it
becomes nearly flat if the  outliers (red and blue) are excluded. This
could  be  an  indication  that  most   non-family  V-types  have  a
different origin with respect to family V-types.\\
The NASA Dawn mission will aid in understanding the weathering process
on Vesta,  thanks to spectral imaging and  global mapping.  Therefore,
the  possibility to  have  detailed spectral  information of  specific
areas  of different  composition and  texture will  shed light  on the
different effects of space weathering on basaltic asteroids.

\acknowledgments

We thank the anonymous referee for the helpful comments on the
manuscript.

\clearpage

\end{document}